%% file: kleinwolt.tex
\documentclass[manuscript]{emulateapj}
\bibpunct{(}{)}{;}{a}{}{,}

\slugcomment{Accepted for publication in ApJ}

\shorttitle{Identification of black hole power spectral components}
\shortauthors{Klein-Wolt \& van der Klis}

\begin{document}

\title{Identification of black hole power spectral components across 
all canonical states}

\author{M. Klein-Wolt \& M. van der Klis}
\affil{Astronomical Institute "Anton Pannekoek" University of Amsterdam, Kruislaan 403, 
1098 SJ Amsterdam, The Netherlands\\}
\email{klein@science.uva.nl}

\begin{abstract}
From a uniform analysis of a large (8.5 Ms) Rossi X-ray Timing Explorer data set of Low Mass X-ray Binaries, we present a complete identification of all the variability components in the power spectra of black holes in their canonical states. It is based on gradual frequency shifts of the components observed between states, and uses a previous identification in the black hole low hard state as a starting point. It is supported by correlations between the frequencies in agreement with those previously found to hold for black hole and neutron stars. Similar variability components are observed in neutron stars and black holes (only the component observed at the highest frequencies is different) which therefore cannot depend on source-specific characteristics such as the magnetic field or surface of the neutron star or spin of the black hole. As the same variability components are also observed across the jet-line the X-ray variability cannot originate from the outer-jet but is most likely produced in either the disk or the corona. We use the identification to directly compare the difference in strength of the black hole and neutron star variability and find these can be attributed to differences in frequency and strength of high frequency features, and do not require the \emph{absence} of any components. Black holes attain their highest frequencies (in the hard-intermediate and very-high states) at a level a factor $\sim$6 below the highest frequencies attained by the corresponding neutron star components, which can be related to the mass difference between the compact objects in these systems.

\end{abstract}




\section{Introduction}
\label{sec:introbhns}

Matter orbiting a compact object in an accretion disk is expected to be relatively unaffected by the nature of the compact object down to a few Schwarzschild radii from the center. Therefore, similar phenomena may occur in the process of accretion onto black holes and low magnetic field neutron stars in low mass X-ray binaries (LMXBs), despite some of the differences between these two types of compact objects, such as the presence or absence of a solid surface or an intrinsic magnetic field. Indeed, many similarities are seen in X-ray observations of neutron stars and black holes, and in fact, traditionally, distinguishing between neutron star and black holes based on other characteristics than X-ray bursts, pulsations and mass estimates has proven to be difficult. However, in both X-ray spectroscopy \citep[e.g.][]{nar97,rut00,barret01,gar01,nar02,dogier03,barret04,titshap05} and X-ray timing, differences between black holes and neutron stars have begun to emerge. In particular with respect to timing, neutron stars are characterized by strong, highly tunable twin kHz quasi periodic oscillations (QPOs) at frequencies up to $\sim1300$ Hz \citep[for a review see][]{klis00,klis06} whose phenomenology is quite different from that of the high-frequency QPOs seen in black holes \cite[e.g.][]{homan03a,homan05a,rem99b,rem02c,stroh01a,klis06}. Also, \cite{sun00} have suggested that in some states neutron stars show considerably more power above several 100 Hz as compared black holes in any state.

\begin{figure}[!htb]
\epsscale{1.0}
\plotone{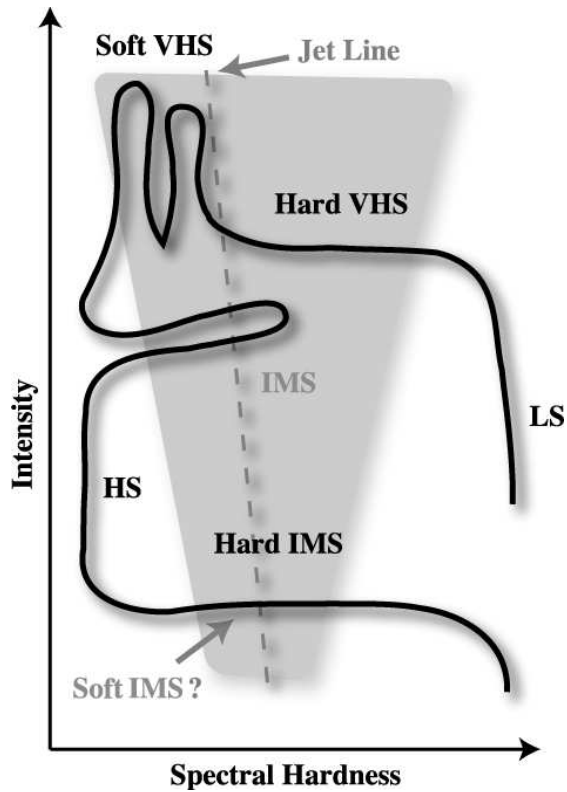}
\caption[]{The canonical black hole states and their location in the hardness-intensity diagram. The track shown is an example of those that are followed by black hole transients in outburst, but persistent sources too can follow parts of such as track. Note that as depicted flares can occur in the HS during which the energy spectrum get harder and the power spectra are very similar to those seen in the hard or soft VHS. In this paper we use the names in bold face, see text. The grey area indicates the IMS region, which we further distinguish in the hard- and soft VHS and the hard IMS. In this figure we also show the jet-line \citep{fender04,kording06} which can be found parallel to the line indicating the transition to and from the HS. The jet-line marks a transition between radio loud to a radio quiet states. Note that during a hard flare in the HS, the radio can be switched on again.\\
}
\label{fig:hidstatesBH}
\end{figure}
The observation of recurrent, correlated spectral and timing behavior led to the introduction of the ''canonical'' states for the black hole and neutron star systems. For the black holes the definition of these states has undergone some evolution over the years. Initially luminosity was seen as the main determinant of state while in later work spectral hardness and two-dimensional classification schemes dominated, and the terminology reflects this historical process \citep[see for instance][]{klis95,tan95,homan01,bell05b,klis06}. We use the state definitions and terminology that resulted from this process, and in particular classify source state primarily based on the relative location in the hardness-intensity diagram (HID) and power spectral characteristics. We distinguish the following states, see Fig.~\ref{fig:hidstatesBH} for a schematic HID: the \emph{low-hard state} (LS), characterized by a hard power-law dominated X-ray spectrum and strong, low-frequency band-limited noise in the power spectrum, the \emph{high-soft state} (HS) with a soft, disk black-body dominated X-ray spectrum and little rapid variability, and an \emph{intermediate state} (IMS), with both power law and disk black-body contributing appreciably to the spectrum and exhibiting the most complex variability characteristics, including most of the QPOs. All three states occur over a wide range in luminosity, although towards the lowest luminosity levels sources consistently enter the LS. Following \cite{bell05b} within the IMS we further distinguish a \emph{hard intermediate state}, which shows somewhat harder X-ray spectra (although not as hard as in the LS) and power spectra with moderately strong band limited noise (BLN; somewhat weaker and at higher frequency than that in the LS), and the \emph{soft intermediate state} with softer X-ray spectra (not as soft as in the HS) and power spectra lacking this BLN component but showing power-law noise. The hard IMS has characteristics in common with the LS and the soft IMS with the HS; the hard--soft IMS transition may be the most physically abrupt one between what may considered to be just two basic black hole states: hard and soft. The IMS was first observed at high luminosity (in the peak of black hole transient outbursts) and for that reason was originally called very high state \citep[VHS;][]{miya91}, and the rapid transitions between hard and soft IMS at high luminosity were already a defining characteristic of the VHS when it was first named. At lower luminosity the IMS usually shows up as hard IMS. To distinguish between high and low luminosity occurrences of the IMS, we shall indicate the high-luminosity realizations of hard and soft IMS as \emph{hard} and \emph{soft VHS}, respectively, while the low flux instances of the IMS will simply be referred to as hard (and if it occurs, soft) IMS. We consider hard IMS and hard VHS as basically the same state but at different luminosity levels, as their power spectra very similar (see for instance Sect.~\ref{sec:lshvhshims}).

For completeness we note that \cite{clintock04} proposed alternative definitions and names for black hole states, predominantly based on the energy spectral behavior and therefore not in one-to-one correspondence with the states distinguished here. They define a 'hard state' which is closest to the LS and a 'thermal-dominant state' closest to the HS, and further distinguish a 'steep power law state' that is closest to the soft IMS. Contrary to the current work, that classification does not aim to provide coverage of all observable states, but rather to exemplify a number of specific combinations of X-ray spectra and rapid X-ray variability in order to provide a focus for physical modeling. As a consequence of this, many observations not fitting the \cite{clintock04} definitions occur. These are described as 'intermediates' between these three states; in particular the hard IMS in their description is intermediate between hard and steep power law states. In the work presented here we will use the canonical state definitions as summarized above, in which the hard and soft VHS and hard IMS are all different realizations of the IMS found in between the LS and HS, as shown in Fig.~\ref{fig:hidstatesBH}. In that figure we also indicated the possibility of soft IMS occurrences during the low luminosity intervals of the IMS, similar to the soft VHS at high luminosities, as suggested for instance by \cite{bell05b}. Up to date no unambiguous detection of such a soft IMS has been reported, perhaps due to a combination of the fact that the source generally moves rapidly through this part of the HID, and the lower statistics at these low luminosities. We finally note that the results presented here are not sensitive to the precise naming scheme for the black hole states.

In Fig.~\ref{fig:hidstatesBH} we also indicate the position of the jet-line which marks the transition between radio-loud and -quiet states \citep{fender04,kording06}. The jet-line in Fig.~\ref{fig:hidstatesBH} is a generic one and in individual cases the transition between radio-loud and -quiet states is not always a sharp one. Radio emission in the radio-quiet (also known as "quenched") states is not ruled out \citep[][and references therein]{kording06} but occurs at flux levels of up to a factor 50 lower than compared to the radio-loud states \citep[][]{corbel01}. Remnant emission from large ejection events that occur close to the right-hand side of the jet-line may still be observed to the left of it \citep[in the radio-quiet states;][]{kording06}. In any case, in individual cases the jet-line is observed to be diagonal (Fig.~\ref{fig:hidstatesBH}), i.e. in a typical transient outburst the transition back to radio-loud at low X-ray luminosity occurs at a spectrally harder position than the quenching of the jet at high X-ray luminosity, but its exact position is still a matter of debate (Fender R., 2007, private communications).

\begin{figure}
\epsscale{1.25}
\plotone{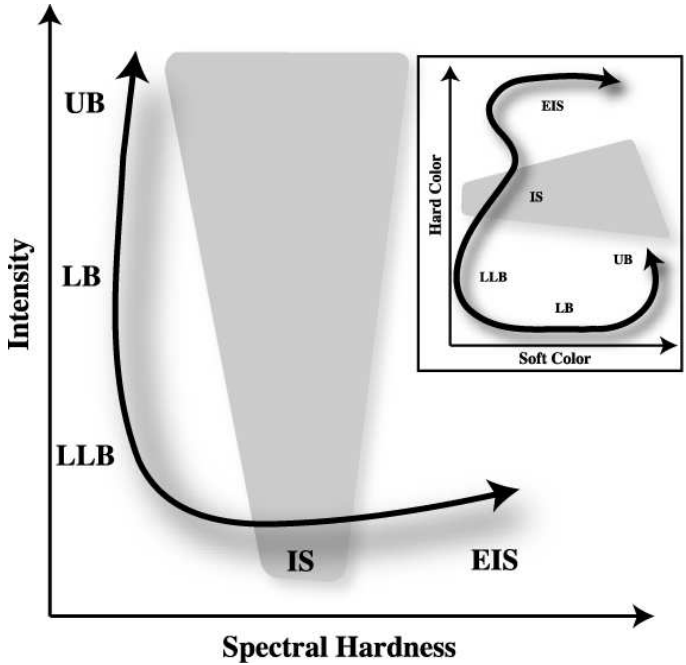}
\caption[]{The canonical atoll neutron star states and their location in the hardness-intensity diagram. Note that we use here the same, but for neutron stars unconventional, black hole HID representation as in Fig.~\ref{fig:hidstatesBH}. The inset shows the same track using the conventional neutron star CD. The track shown is a typical one followed for neutron star atoll sources (persistent or transient); some sources show transitions between all states, while others are only found in few of them.}
\label{fig:hidstatesNS}
\end{figure}

The neutron star LMXBs are subdivided into Z sources and atoll sources based on differences in their correlated X-ray spectral/timing behavior. The names are derived from the tracks these sources produce in a color-color diagram \citep[CD;][]{has89}. Below we will only discuss the atoll sources, which have the following states (Fig.~\ref{fig:hidstatesNS}): the extreme island state (EIS), the island state (IS) and the banana branch subdivided into lower left banana (LLB), lower banana (LB) and upper banana (UB) \citep[for a review see for instance][]{klis06}. The precise topology of these tracks at low X-ray luminosity (EIS) is a matter of some debate \citep{gier02,muno02,barret02,straaten03,reig04} but that is of no concern to us here.

The behavior of the black holes and the neutron stars in their different states shows similarities. Based on the spectral and timing characteristics, \cite{klis94a} already suggested that the spectrally hard black hole state (LS) is similar to the spectrally hard neutron star atoll state (EIS; although at that time the distinction between IS and EIS was not yet commonly made), and that some of the softer black hole and neutron star states might also be identified with one another. Another way to associate neutron star and black hole phenomenology is by using the strong correlations that exist between the characteristic frequencies of most of their power spectral components. Some of these frequency-frequency relations look the same for black holes and (atoll) neutron stars, suggesting that the same physical processes underly these power spectral components, or, at least, the correlations between them. The frequency-frequency relations first identified were the Wijnands-van der Klis  relation \citep[WK;][]{wvdk99} and the Psaltis-Belloni-van der Klis relation \citep*[PBK;][]{pbk99}. The WK relation relates the break frequency of the band-limited noise and the centroid frequency of a low-frequency QPO above this break, and is the same for atoll sources and black holes. Z sources follow a parallel but slightly offset relation. The PBK relation relates the frequency of the 1-10 Hz high-frequency bump and that of a 0.1--1 Hz low-frequency QPO in the black hole and weak neutron star sources, and coincides with an extrapolation of the relation between the 100-1000 Hz lower kilohertz QPO frequency and the 10--100 Hz low-frequency (LF) QPO frequency found in both atoll and Z sources. More recent studies confirm these results, although it is also clear that the entire phenomenology cannot be caught in just these two correlations \citep[e.g.][]{nowak00,bpk02,pott03,kalemci03,bell05b,stratos06}. \cite{straaten02,straaten03} for the atoll sources extended this set of relations into a proposed 'universal scheme' of correlations which encompass the WK and PBK relations among several others.  \cite{war02} and \cite{mauche02} show that the PBK relation may even extend to white dwarf systems, which implies that even while some of the neutron star and black hole frequencies may require strong field gravity for their generation, at least the mechanism producing the {\it correlations} between the frequencies must be generic to a broad class of accretion flows.

A complication regarding the frequency-frequency relations is the proper identification of the different types of variability phenomena, especially in the black hole power spectra. For the (neutron star) atoll sources the identification of the different components in the power spectra now seems relatively well established \citep{disalvo01,straaten02,straaten03,straaten05,diego05,manu05}, even across different states, as components are observed to gradually change across state transitions. For the black holes no complete set of characteristic variability components observed across all canonical states exists. Only in the case of the black hole LS an identification, similar to that of the neutron star EIS, has been suggested \citep{wvdk99,pbk99,nowak00,bpk02}.

In this paper we attempt for the first time to perform a complete identification of all black hole variability components across all canonical states. For this purpose we use the identification of the variability components in the black hole LS as presented by \cite{bpk02}, and the observed amplitude changes and shifts in frequency of these components during transitions to the other states as observed in a number of black hole transients. Using these new identifications, we study the frequency-frequency relations and the variations in amplitude with frequency in a large number of black hole observations. We find that the frequency-frequency relations confirm our identifications, and are able to identify which components are responsible for the differences between the black hole and neutron star power spectra.

The analysis presented here is part of a much larger program in which we compare all of the available public \emph{Rossi X-ray Timing Explorer (RXTE) Proportional Counter Array (PCA)} data for the eight neutron star sources Aquila~X-1, 4U~0614+091, Terzan 2, 4U~1636--536, 4U~1728--34, 4U~1608--52, GX~3+1, 4U~1820--30 and the nine black hole sources XTE~J1118+480, GX~339--4, XTE~J1650--500, GRO~J1655--40,XTE~J1748--288, GRS~1758--258, XTE~J1859+266, XTE~J1550--564 and 4U~1543--47. Our current analysis covers total exposure of $\sim8500$ ks up to AO8.

\section{Data analysis}
\label{sec:analysis}

\begin{figure}[!htb]
\epsscale{1.0}
\plotone{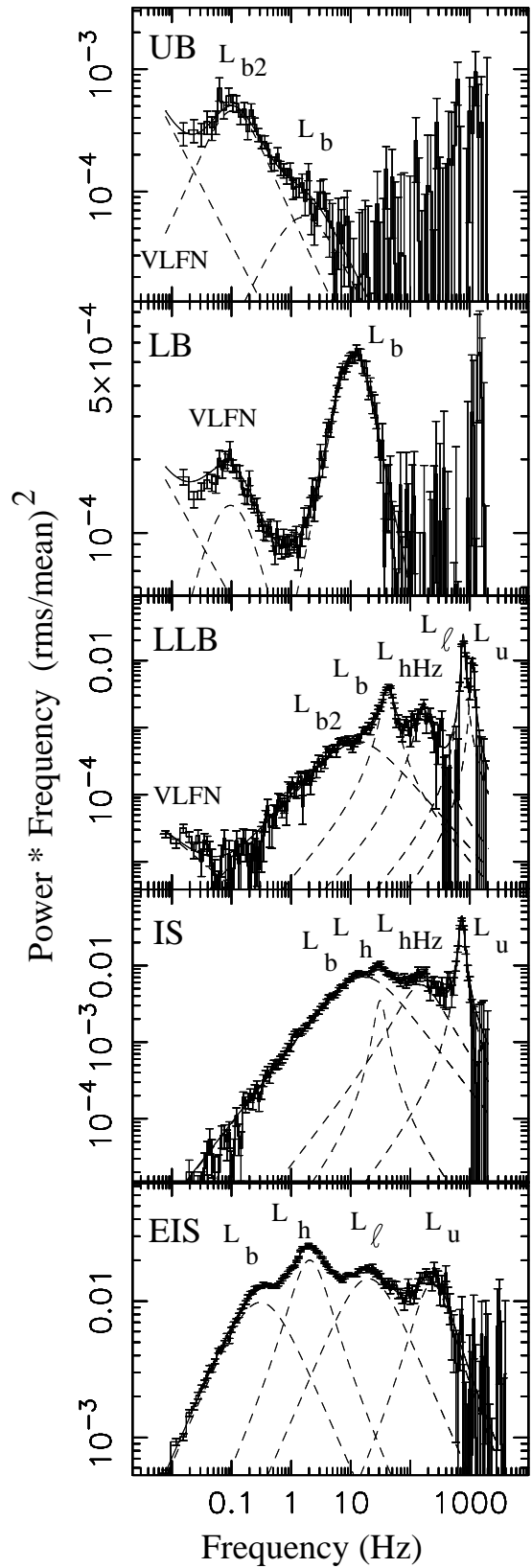}
\caption[]{The identification of the power spectral components for 
the different neutron star atoll states.  The EIS power spectrum is a
characteristic one from 4U~0614+091, the IS and LLB ones are from
4U~1728--34 and the LB and UB ones are from GX~9+1 \citep{reer04}. }
\label{fig:reeks}
\end{figure}

We analyze the data in a largely automated fashion, which allowes for a consistent analysis of our large data set. The data are comprised of 2860 observations, each identified by a unique \emph{RXTE} observation ID and consisting of one to three contiguous, typically 1--3 ks intervals separated by Earth occultations and/or South Atlantic Anomaly (SAA) passages. Before performing the analysis, any bursts and dropouts are removed automatically using 16 second time-resolution light curves created from \emph{Standard 2} data. For bursts we locate the first point whose count rate exceeds twice the average rate of the observation (this does not have to be the maximum of the burst) and trace the start and end of the burst by finding the points around this point where the count rate is again similar (within 1$\sigma$) to the average flux. Furthermore, we remove weak observations for which the source was not detected above $10\sigma$. This results in the removal of data with a low intensity level ($<10^{-3}$ Crab), which comprises only a few percent of the total number of observations for each source.

We use all available high time resolution data to create Fourier power spectra. For high count rate sources the lower energy channels 0--8 or 0--13 are excluded from the data in order to exclude instrumental effects, as suggested by many authors \citep[e.g.][and see Klein-Wolt et al. 2005]{sun00}. We construct Leahy normalized power spectra using 128 s data segments and 1/8192 s time bins such that the lowest available frequency is 1/128 Hz and the Nyquist frequency is 4096 Hz. These power spectra are averaged per observation (see above) and the resulting average power spectra are converted to source fractional rms squared per Hz. We subtract the deadtime modified Poisson noise using the method of \cite{klein05} based on the expression of \citet{zhang95}, shifting the Poisson function by a small amount ($<1\%$) to match the power in a high frequency range where no source power is expected. For most of the observations the range between 2 and 4 kHz is used for this. 

The issue of correcting for deadtime effects in temporal analysis of RXTE data is much debated and still unresolved \citep[see for instance][]{vik94,zhangetal95,zhang95,klein05,stratos06}. However, in the present work the effects are regarded as negligible as the power spectra above $\sim100$ Hz depend only slightly on the way the Poisson level is subtracted. As shown in \cite{klein05} the systematic uncertainty due to the residual Poisson noise in the rms amplitude of a broad high frequency feature when using the scaled Zhang function is in the order of $\sim1$--$10$\%. For a small fraction ($\sim$5\%) of the observations the 2--4 kHz power spectrum shows a increasing or decreasing trend that did not match the predicted shape of the Poisson function. In those cases a lower range, between a few 100 Hz and $\sim$1000 Hz is used to match the Poisson function, and the power spectral information above $\sim100$ Hz is ignored. 

We fit a characteristic subset (see Sect.~\ref{sec:ident}) of 160 observations using a multi-Lorentzian model \citep[cf.][]{bpk02}. We describe the Lorentzians by their characteristic frequency (the frequency where the component contributes most of its variance per logarithmic frequency interval) $\nu_{\rm max}=\sqrt{\nu_0^2 + \Delta^2}$, where $\nu_0$ is the centroid frequency and $\Delta$ the HWHM of the Lorentzian, and by their inverse relative width (coherence) $Q$, defined as $\nu_0/2\Delta$. We only keep those components in the fits whose significance, based on the error in the integrated power (from 0 to $\infty$), is more than 3$\sigma$ or whose inclusion gives a $>3\sigma$ improvement of the fit according to an F-test. Narrow Lorentzians, with Q larger than 2, are referred to as quasi periodic oscillations (QPOs), broad or BLN features have Q$<2$. For some of these features $\nu_0$, and hence $Q$, is not well constrained and runs away to frequencies below zero. In fitting the power spectrum these features are fixed at $\nu_0=0~(Q=0)$, with little effect on the overall quality of the fit. For plotting the power spectra we use the power times frequency representation ($\nu{\rm P}_{\nu}$), where each power spectral density estimate ${\rm P}_{\nu}$ is multiplied by its Fourier frequency $\nu$. For a Lorentzian this representation helps to visualize the characteristic frequency $\nu_{\rm max}$, as in $\nu{\rm P}_{\nu}$ the Lorentzian's maximum occurs at $\nu_{\rm max}$.

\begin{figure}
\epsscale{1.0}
\plotone{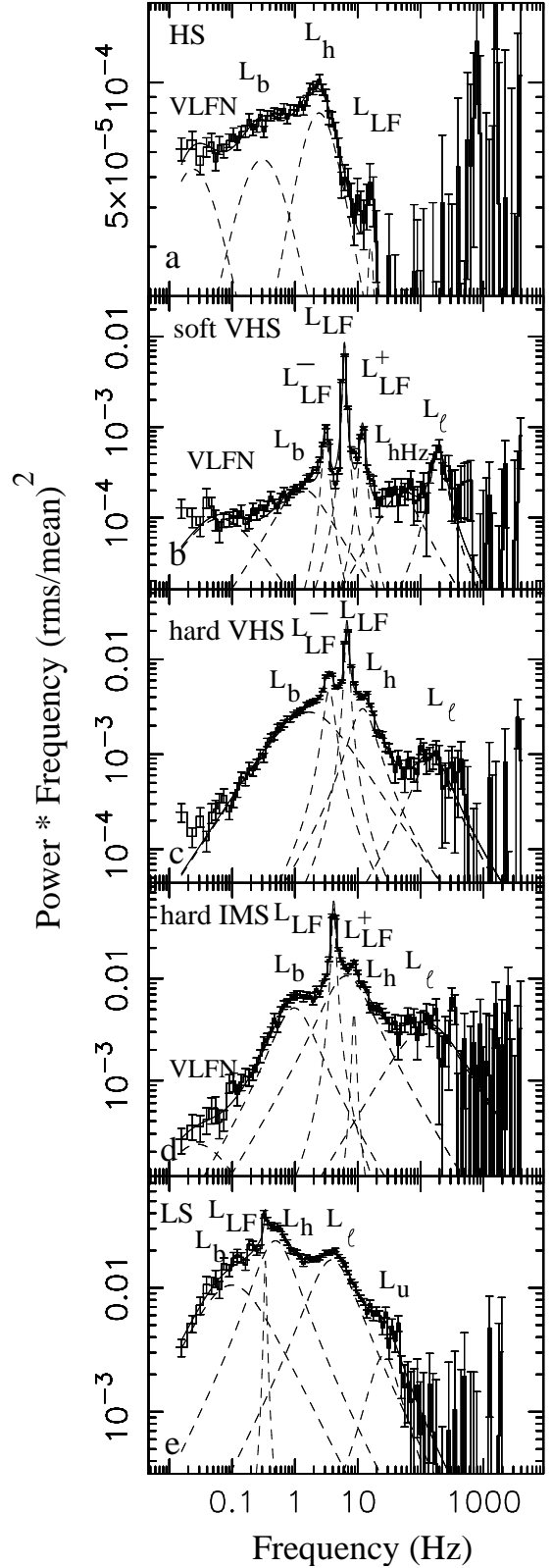}
\caption[]{The characteristic individual power spectra for all the black hole states. In each panel we identified the different components in the power spectra as explained in the text. Panels a--d show the power spectrum of XTE~J1550-564 in the HS (obsID 40401-01-15-00), hard VHS (obsID 40401-01-57-00), soft VHS (obsID 40401-01-55-00) and the hard IMS (obsID 50135-01-01-00), and in panel e we show the LS power spectrum of GX~339--4 (obsID 20181-01-01-00). Note that in panel b an extra VLFN component was added to fit a weak noise component at low frequencies ($\sim0.01$ Hz).\label{fig:identify}}
\end{figure}

\section{Identification of the components in the black hole power spectra.}
\label{sec:ident}

\begin{figure}
\epsscale{0.9}
\plotone{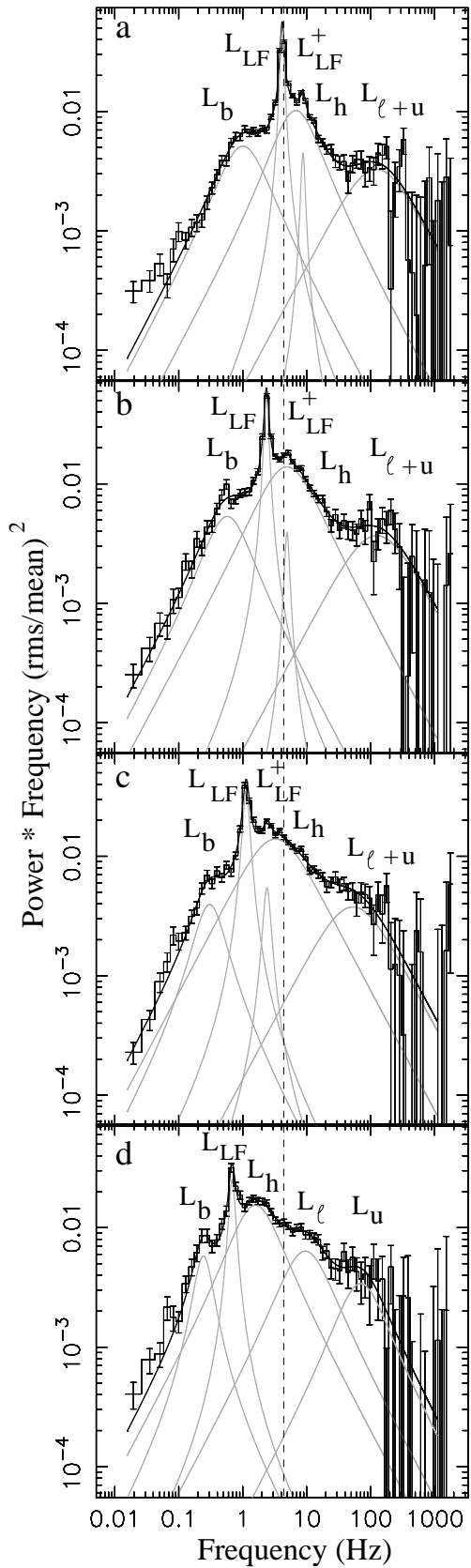}
\caption[]{The transition from a typical hard IMS to a typical LS in 
XTE~J1550-564, from MJD 51676.4 to 51686.3 (observations 50135-01-01-00, 50135-01-02-00, 50135-01-03-00 and 50135-01-05-00 in panels a, b, c and d respectively). The components in the power spectra are identified using the LS identification by \cite{bpk02} and then tracing back the components as they change frequency. The dashed vertical line indicates the frequency of the L$_{LF}$ component in the first power spectrum (panel a).\label{fig:overgang}}
\end{figure}

\begin{figure}
\epsscale{1.0}
\plotone{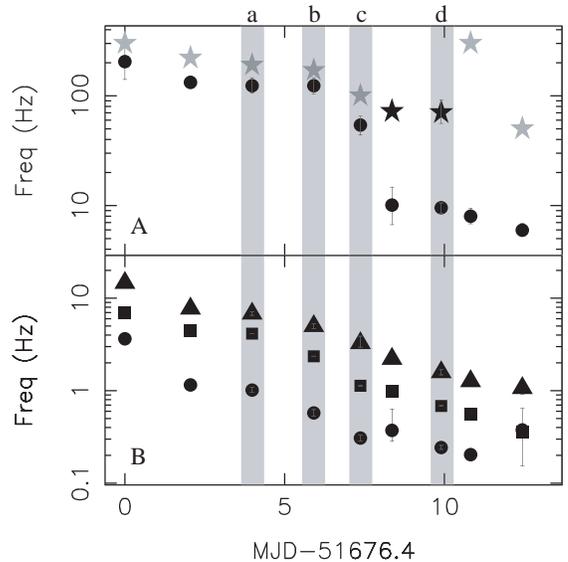}
\caption[]{The change in frequency of the components during the state changes of XTE~J1550-564. Panel A shows L$_{\ell}$ (bullets) and L$_{u}$ (stars) and panels B shows L$_{b}$ (bullets), L$_{LF}$ (squares) and L$_{h}$ (triangles). Note that in panel A the gray stars give the upper limits for L$_{u}$ in some observations for which L$_{u}$ was not significantly detected. At the top of the panels the letters a--d correspond to the observations shown in the corresponding panels in Fig.~\ref{fig:overgang}.}
\label{fig:afname1550}
\end{figure}

In this section we present an identification for the components detected in the black hole power spectra in all the canonical states. It is based on the identification by \cite{bpk02} for the LS power spectra which makes use of the similarity to the neutron star EIS, and uses the observed, often gradual transitions between the states to extend the identification to all black hole states. These transitions allow us to follow the components in the power spectra as they change both in strength and frequency. For the identification of the states, in addition to the power spectral shape and its changes, we also use the instantaneous position of each source relative to its track in the color-color and hardness-intensity diagrams \citep[e.g.][]{has89,homan01,bell05b}. After inspection of all power spectra obtained from the large set of observations of the eight neutron star and nine black hole systems listed in Sect.~\ref{sec:analysis} we selected 160 characteristic individual power spectra covering all states such that all the different recurrent spectra of each state are represented. These characteristic power spectra are fitted using the method explained in Sect.~\ref{sec:analysis}. It is important to emphasize that, as a direct result of the selection of characteristic power spectra, anomalous power spectra or features in power spectra are poorly sampled. As we will explain below and address further in Sect.~\ref{sec:caveats}, this is especially true for the high frequency QPOs occasionally found in black hole power spectra in addition to the broad high frequency features \citep[e.g.][]{stroh01a,rem02c,homan05a,homan03a,miller01}; more detailed analysis is necessary to investigate their behavior in full, but this is beyond the scope of this paper.

For reference, in Fig.~\ref{fig:reeks} we show the power spectra that in our analysis we found to be typical for the different neutron star atoll states. The identification of the power spectral components in Fig.~\ref{fig:reeks} follows \cite{straaten03}. In order of increasing frequency we distinguish the following components: VLFN,
L$_{b2}$, L$_{b}$, L$_{h}$, L$_{hHz}$, L$_{\ell}$ and L$_{u}$. Note that,
\cite{straaten05} regard the band limited noise component found in the EIS in between L$_{h}$ and L$_{u}$ as possibly distinct from L$_{\ell}$ found at higher frequency in other neutron star states, and hence designate it as L$_{\ell ow}$ (L$_{\ell}$ and L$_{\ell ow}$ do not occur together). For simplicity we use the designation L$_{\ell}$ for both of them.

\subsection{LS, hard VHS \& hard IMS}
\label{sec:lshvhshims}

In Fig.~\ref{fig:identify}e we show the identification of the components in a typical black hole LS power spectrum for GX~339--4. We use the identification as presented by \cite{bpk02} for LS power spectra of XTE~J1118+480, but it is necessary to add an extra L$_{h}$ component analogous to that seen in neutron stars. This component was not present in the XTE~J1118+480 data, but is clearly detected in GX~339--4 (Fig.~\ref{fig:identify}e). Thus, our identification very closely matches that of the EIS power spectra of the neutron stars 1E~1724--3045 and GS~1826--24 presented by \cite{bpk02} and also matches the EIS presented in Fig.~\ref{fig:reeks}. As also noted by \cite{bpk02} the black hole LS power spectrum differs from the neutron star EIS one by the presence of a narrow QPO (L$_{LF}$) near L$_{h}$, and by less power at frequencies above $\sim100$ Hz.

\subsubsection{Hard IMS to LS transition at low flux}
\label{sec:imsls}

\begin{figure}
\epsscale{0.85}
\plotone{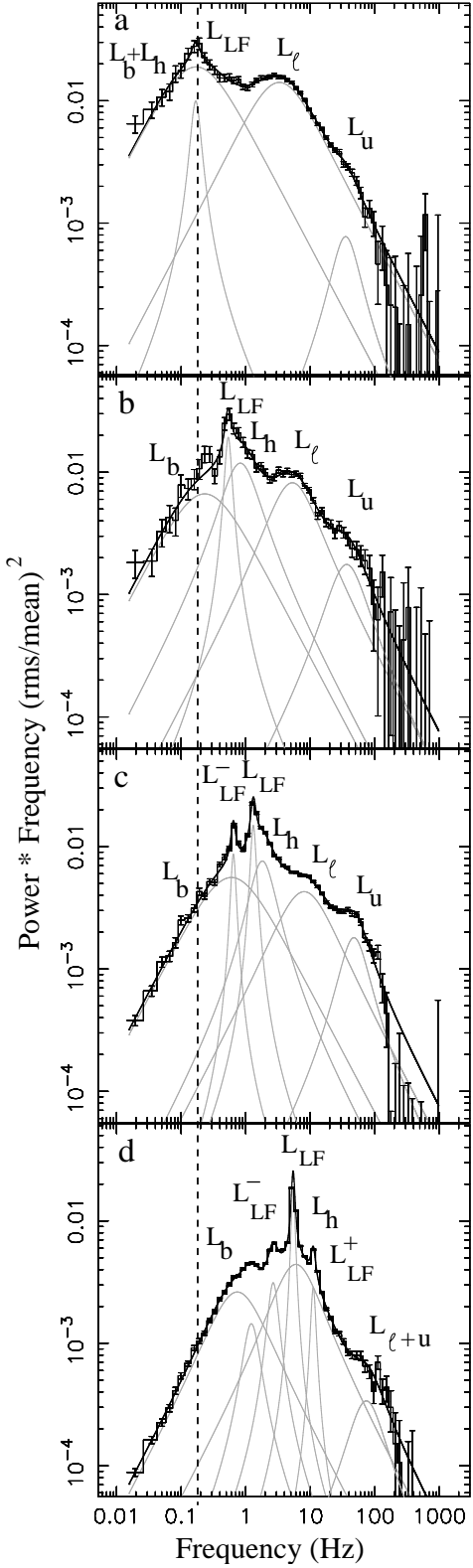}
\caption[]{The transition from a typical LS to a typical hard VHS in GX~339--4, from MJD 52377.09 to 552405.71 (observations 40031-03-02-04, 70110-01-09-00, 70108-03-01-00 and 70109-04-01-01 in panels a, b, c and d respectively). The components in the power spectra are identified using the LS identification by \cite{bpk02} and then tracing back the components as they change frequency. The dashed vertical line indicates the frequency of the L$_{LF}$ component in the first power spectrum (panel a). Note there is some extra structure in the BLN component in panel d: it is detected at 1.23 Hz at a weak level of $\sim3\sigma$ (single trial significance). From the frequency of this component we conclude that it is most likely not an harmonic of the L$_{LF}$ component. Additional structure in the BLN is only found sporadically in the hard- and soft VHS.\label{fig:overgang2}}
\end{figure}

\begin{figure}
\epsscale{1.0}
\plotone{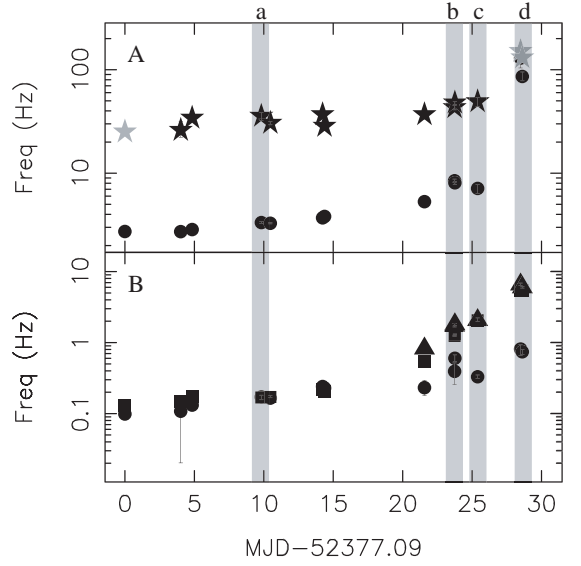}
\caption[]{The change in frequency of the components during the state changes of GX~339--4. Panel A shows L$_{\ell}$ (bullets) and L$_{u}$ (stars) and panels B shows L$_{b}$ (bullets), L$_{LF}$ (squares) and L$_{h}$ (triangles). Note that in panel A the gray stars give the upper limits for L$_{u}$ in some observations for which L$_{u}$ was not significantly detected. At the top of the panels the letters a--d correspond to the observations shown in the corresponding panels in Fig.~\ref{fig:overgang2}.}
\label{fig:afname339}
\end{figure}

In order to extend this identification to the other black hole states we follow in detail the transitions between the states of the black hole sources in our sample. For most of the sources we find that the transitions at high flux between the LS and hard VHS and those at low flux between the hard IMS and the LS are sufficiently gradual to allow us to track the components in the power spectrum from observation to observation, as they change in frequency and strength. 

As an example of such a gradual transition we show in Fig.~\ref{fig:overgang} the transition from the hard IMS to the LS (top to bottom) during the decay of the 2000 outburst of XTE~J1550--564 as a sequence of four individual observations covering twelve days (MJD 51676.4 -- 51686.3). The total sequence consists of nine observations, some of which show only small changes in frequency. Figure~\ref{fig:afname1550} shows the frequency measurements for all the components in those observations. Application of the identification of the components in the LS to Fig.~\ref{fig:overgang}d and tracing the components back towards the hard IMS leads to the identifications shown in Figs.~\ref{fig:identify}e and \ref{fig:overgang}a--d, as follows. The sequence in Fig.~\ref{fig:overgang} shows that the strongest low frequency QPO, designated L$_{LF}$, gradually shifts in frequency from $\sim7$ Hz in the hard IMS to $\sim0.5$ Hz in the LS. We conclude that this QPO in the hard IMS power spectra is the same feature (but moved to higher frequencies) as that in the LS (Fig.~\ref{fig:overgang}d), and hence we identify it as such. We designate the weaker LF QPO, which is approximately a factor two in frequency above L$_{LF}$, as L$_{LF}^+$ (Figs.~\ref{fig:overgang}a, b, and c). The identification of the other, broader (noise) components in the low frequency part of the power spectrum (below several tens of Hz) derives from the strong L$_{LF}$ QPO and is based on the LS identification: for all states we identify the broad feature below L$_{LF}$ as L$_{b}$ and the one above it as L$_{h}$ (Figs.~\ref{fig:identify} and \ref{fig:overgang}). Note that L$_{LF}$ and L$_{h}$ are often close in frequency (separated by not more than a few Hz), and seem to form a closely coupled set of features. Figure~\ref{fig:afname1550}b shows all our measurements of these three frequencies; their common gradual decrease over time is obvious.

The situation is more complicated at higher frequencies. Only on two occasions we detect simultaneously two significant components; one of these power spectra is plotted Fig.~\ref{fig:overgang}d and the frequencies of both cases can be seen in Fig.\ref{fig:afname1550}a (black symbols). In those cases we designate the component closest in frequency to L$_{LF}$ and L$_{h}$ as L$_{\ell}$ and the higher frequency one as L$_{u}$ in analogy to other LS (and EIS) power spectra (Fig.~\ref{fig:identify}e). In the event only one feature is detected we identify it as L$_{\ell}$. However, we cannot exclude that in those cases we see a blend of L$_{\ell}$ and L$_{u}$. An argument in favor of this is that the single L$_{\ell}$ features are broader than those that are accompanied by an L$_{u}$ component (average Q of $\sim0.3$ compared to $\sim0.02$). 

For the cases when only one high frequency feature is significantly detected Fig.~\ref{fig:afname1550}a also shows the frequencies at which the highest value for the 95\% upper limit (using Q=0.9 and a frequency range of 100--1000 Hz, comparable to the other cases when L$_{u}$ was detected) on the rms amplitude for a second high frequency feature is reached (grey symbols). These upper limits are relatively high, between 3 and 10 \% rms, which allows the presence of L$_{u}$ in those cases at a similar strength as the detections in other observations. As can be seen in Fig.~\ref{fig:afname1550}a these upper limit values suggest that L$_{\ell}$ and L$_{u}$ are close in frequency (i.e. there is no evidence for two separate components) until $\nu_{\ell}$ drops to $\sim10$ Hz (around day 8) and the frequency difference between $\nu_{\ell}$ and $\nu_{u}$ becomes much larger. After the two LS power spectra where L$_{\ell}$ and L$_{u}$ are detected separately on days 8 and 10, the next two spectra on days 11 and 13 admit an additional L$_{u}$ component well above L$_{\ell}$; this component is not, however, significantly detected. Figures~\ref{fig:overgang} and \ref{fig:afname1550} show all the components L$_{b}$, L$_{LF}$, L$_{h}$ and L$_{u}$ gradually decreasing in frequency as the source moves from the hard IMS (Fig.~\ref{fig:overgang}a) to the LS (Fig.~\ref{fig:overgang}d). Only L$_{\ell}$ shows a sudden drop in frequency (near day 7). This is due to the choice to designate single high frequency components all as L$_{\ell}$, which was made in order to remain consistent with \cite{pbk99}, see also Sect.~\ref{sec:caveats}.

\subsubsection{LS to hard VHS transition at high flux}

In Fig.~\ref{fig:overgang2} we show a similarly gradual transition from a typical LS to a typical hard VHS (top to bottom) during the rise of the 2002 outburst of the black hole transient GX~339--4. The total sequence lasts from MJD 52377.09 to 52405.71 and consists of 13 observations of which 4 are shown in Fig.~\ref{fig:overgang2}. All frequency measurements are plotted in Figs.~\ref{fig:afname339}. Again, the components can be identified as explained above using the LS power spectrum as a starting point, and tracing the L$_{LF}$ QPO as it changes in frequency. Analogously to the previous sequence, we identify the strongest LF QPO as L$_{LF}$ and call the weaker LF QPOs a factor two above and below it L$_{LF}^+$ and L$_{LF}^-$, respectively. The initial LS power spectrum in Fig.~\ref{fig:overgang2}a (corresponding to the first 6 observations of the transition, see Figs.~\ref{fig:afname339}) is somewhat different from the LS spectrum in Fig.~\ref{fig:overgang}d. In Fig.~\ref{fig:overgang2}a, the features are found at lower frequencies compared to those in Fig.~\ref{fig:overgang}d, and only one broad low frequency noise component is detected in addition to L$_{LF}$. We interpret this component as a blend of the L$_{b}$ and L$_{h}$ components, which here are sufficiently close together to appear as one feature. This interpretation is supported by the observation that as the source moves from the LS towards the hard VHS the individual L$_{b}$ and L$_{h}$ components both become visible and gradually become sharper. The L$_{h}$ component increases in frequency at a higher rate than L$_{b}$ (the relative change in frequency of L$_{h}$ is $\sim 0.5 $ per day, as compared to $\sim 0.4 $ per day for L$_{b}$), which also helps the two components separate. After the frequencies have increased somewhat, in Fig.~\ref{fig:overgang2}b (day 21) we see a LS power spectrum that is quite similar to that in Fig.~\ref{fig:overgang}d, despite the fact that Fig.~\ref{fig:overgang} depicts a low luminosity transition in XTE~J1550-564 and Fig.~\ref{fig:overgang2} a high luminosity one in GX~339--4. The frequencies of L$_{b}$, L$_{LF}$ and L$_{h}$ gradually increase there after, with $\nu_{LF}$ and $\nu_{h}$ always very close (Fig.~\ref{fig:afname339}b).

At higher frequencies for most of the observations in this sequence we significantly detect both L$_{\ell}$ and L$_{u}$ which change together with gradually decreasing frequencies until day 25 (see Fig.~\ref{fig:afname339}a). After that we detect only one significant feature, which again for reasons given in Sect.~\ref{sec:imsls} and further discussed in Sect.~\ref{sec:caveats}, we designate L$_{\ell}$. This feature is found at higher frequencies compared to the L$_{\ell}$ components we detect before day 25 and, like in the case of XTE~J1550--564, we may be seeing a blend here of L$_{\ell}$ and L$_{u}$. Note that again the upper limit values for L$_{u}$ components in those cases are close in frequency to L$_{\ell}$. 

So, up to day 25 all the components gradually change in frequency (see Fig.~\ref{fig:afname339}) and strength causing the general power spectral characteristics to change from LS to hard VHS, and around day 28 probable L$_{\ell}$-L$_{u}$ blends occur.

In total we inspected $14$ of such transitions between the LS and either the hard VHS or hard IMS in our data set and all were similar to the two cases reported here. The recent outbursts of GX~339--4 and GRO~J1655--40 also confirm our results. The very similar LS power spectra at both high and low luminosity together with the large similarity between the power spectra of the hard VHS and hard IMS (Fig.~\ref{fig:reeks}), and the gradual changes in frequency of most components in the power spectra during state changes allows us to identify the components in the hard VHS and hard IMS as presented in Fig.~\ref{fig:identify}c and d. We note that the sudden jump in $\nu_{\ell}$ that is not seen in the other frequency components could be avoided by identifying all high frequency features in these two transitions above $\sim20$ Hz as L$_{u}$ -- as discussed in Sect.~\ref{sec:caveats} the choise was made to remain consistent with \cite{pbk99}.

\begin{figure}
\epsscale{1.0}
\plotone{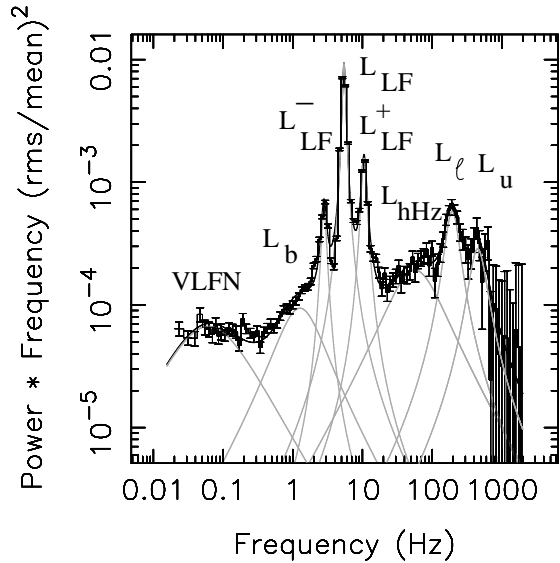}
\caption[]{An example of an observation (observation ID 30191-01-33-00, XTE~J1550-456) with L$_{hHz}$, L$_{\ell}$ and L$_{u}$. We detect these features at 57$\pm5$ Hz (10$\sigma$, 2.1\% rms), 194$\pm4$ Hz (7$\sigma$, 2.0\% rms) and 451$\pm28$ Hz (4.1$\sigma$, 1.6\% rms). }
\label{fig:3hfqpos}
\end{figure}

\subsection{Soft VHS}
\label{sec:softvhs}

The transitions between the soft VHS and either the hard VHS or the HS, are less gradual. However, from our fits we find that in principle the soft VHS power spectra resemble those of the hard VHS, with similar QPO structure and the same, but much weaker broad noise components \citep[see Fig.~\ref{fig:identify}b and c and for instance][for examples of soft and hard VHS spectra]{miya91,taki97,rem02a,cas04}. It is the abrupt change in the strength of the noise components (which occasionally coincides with a shift in QPO frequencies in the order of a few Hz) that mostly defines the sudden change in the overall power spectrum during the hard--soft VHS transitions already reported by \citet{miya91}. This allows us to identify the structure of QPOs around 6 Hz in the soft VHS with those QPOs found at similar (but occasionally slightly higher, in the order of a few Hz) frequencies in the hard VHS. Again we identify the strongest LF QPO as L$_{LF}$, and those a factor $\sim2.0$ above and below this as L$_{LF}^+$ and L$_{LF}^-$. At frequencies below the L$_{LF}$ QPOs the soft VHS power spectra have a weak VLFN (at similar strength and frequency occasionally found in the hard IMS, see Fig.~\ref{fig:identify}d) and L$_{b}$ component (Fig.~\ref{fig:identify}b), while a broad component similar to L$_{h}$ (such as is seen in the hard VHS and hard IMS) is \emph{not} detected in the soft VHS power spectra. At the highest frequencies we can find up to three significant features, see for instance Fig.~\ref{fig:3hfqpos}, which we designate in order of increasing frequency: L$_{hHz}$, a relatively broad feature found above the L$_{LF}$ QPOs, L$_\ell$ (which has a somewhat higher $Q$ than in the hard VHS and hard IMS), and above that, on some occasions, L$_{u}$. The reason for the separate designation in L$_{hHz}$ is that it seems to be a different component than other relatively broad features like L$_h$ and L$_{\ell}$ that are close in frequency. This is further discussed in Sect.~\ref{sec:caveats}.

\subsection{HS}

In most of the sources in our sample the transitions to and from the HS are associated with sudden (in the order of days) changes in the power spectra where the overall power spectral shape changes drastically compared to that in the soft IMS, hard VHS or hard IMS. In some HS observations no significant features are detected in the power spectra, most likely due to a combination of weak (intrinsic) power and relatively low count rates. In those cases averaging more power spectra (i.e. combining more than one observation) produces power spectra similar to those individual HS observations that do show significant power spectral features (Fig.~\ref{fig:identify}a). However, in either case the sudden transitions to qualitatively different power spectral shapes prevent the individual components in the power spectra from being traced during transitions to and from the HS. Hence, similar identification methods as used above can not be applied. Instead, we attempt to identify the components in the HS based on their various intrinsic characteristics, in particular their width and their relative frequency with respect to other features in the power spectrum. No components with frequencies above $\sim$20 Hz are detected similar to L$_{hHz}$, L$_\ell$ or L$_u$. Similarly to the identification in LS power spectra, we refer to the broad noise components as VLFN, L$_{b}$ and L$_{h}$ and identify the QPOs with L$_{LF}$ components (Fig.~\ref{fig:identify}a). This results in a good match to the frequency-frequency relations presented below. As the identification in the HS is not based on tracing components through the state transitions, we consider them as less secure.

\section{Frequency-frequency relations}
\label{sec:ffrel}

\begin{figure}
\epsscale{1.2}
\plotone{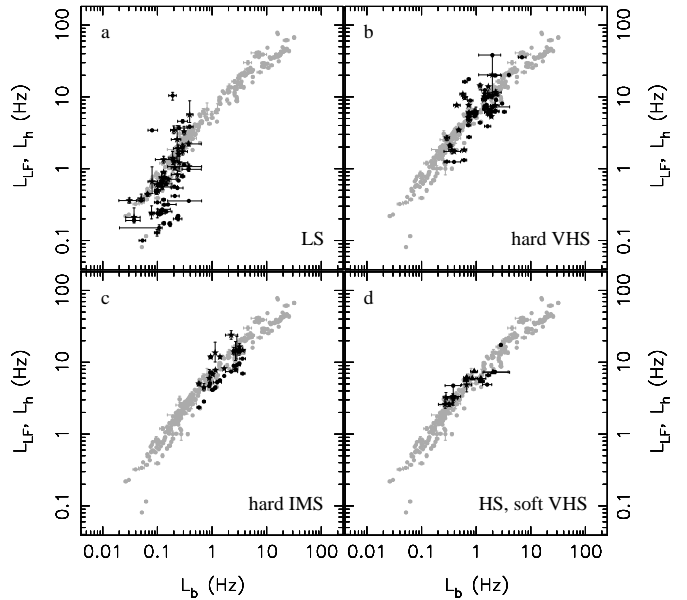}
\caption[]{In panels a--d we show the WK relation (gray bullets) together with our measurements for the relation between the frequencies of L$_{b}$ and L$_{LF}$ (black bullets) and between L$_{b}$ and L$_{h}$ (black stars) for the LS, hard VHS, hard IMS and the HS. In panel d we also show our measurements for the relation between the frequencies of L$_{b}$ and L$_{LF}$ (black triangles) for the soft VHS. The gray points are the original points from the WK relation \cite{wvdk99} complemented by points from \cite{straaten02,straaten03,straaten05} and \cite{bpk02}. \label{fig:wkrel}}
\end{figure}

\begin{figure}
\epsscale{1.2}
\plotone{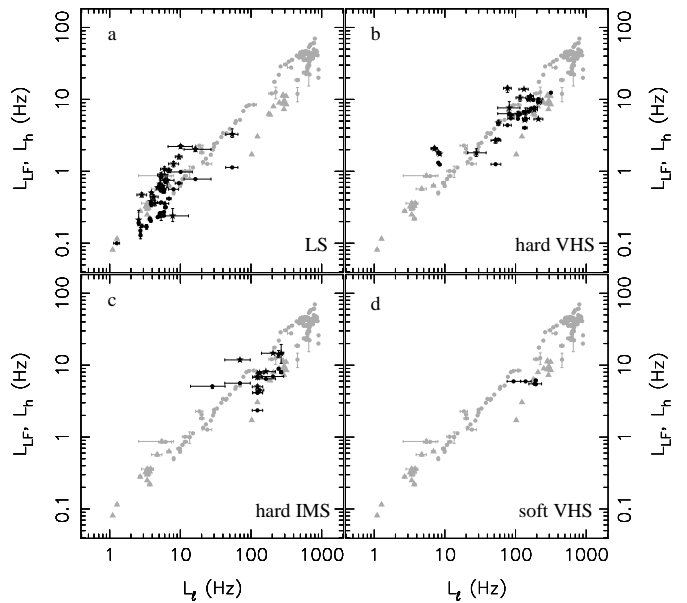}
\caption[]{In panels a--d we show the PBK relation (gray bullets for the neutron star points and gray triangles for the black hole points) together with our measurements for the relation between the frequencies of L$_{\ell}$ and L$_{LF}$ (black bullets) and that between L$_{\ell}$ and L$_{h}$ (black stars) for the LS, hard VHS, hard IMS, soft VHS and the HS. The gray points are taken from \cite{bpk02}.\label{fig:pbkrel}}
\end{figure}

If the components we identified in the hard and soft VHS, hard IMS and HS using the methods outlined in Sect.~\ref{sec:ident} are the same as those found at lower frequencies in the LS, then the frequency-frequency correlations previously identified between pairs of components in (mostly) the LS might be expected to hold across the different canonical states. We fit the power spectra of 160 characteristic observations of our nine black hole sources (50 LS, 50 hard VHS, 25 hard IMS, 10 soft VHS and 25 HS) and identify the components following the scheme outlined in Sect.~\ref{sec:ident}. In each of the panels in Fig.~\ref{fig:wkrel} we show in gray the original WK data \citep[][which include black holes and neutron stars]{wvdk99} complemented with points taken from \cite{straaten02,straaten03,straaten05} and \cite{bpk02}. All points were obtained from fits using only Lorentzians, except the original WK points which were obtained using a broken power law fit and $\nu_{0}$ values (see also Sect.~\ref{sec:analysis}). Together, these points form a narrow band in the diagram, suggesting that the differences between these two different representations are unimportant as also concluded by \cite{bpk02}. Likewise, the PBK \citep[][]{pbk99} relation is shown in gray in each of the panels in Fig.~\ref{fig:pbkrel}; for this we used the $\nu_{max}$ values from \cite{bpk02}.

In black we plot in Figs.~\ref{fig:wkrel} and \ref{fig:pbkrel} the results of our fits to the
characteristic individual power spectra in the different canonical black hole states. As mentioned before, L$_{LF}$ and L$_{h}$ are similar in frequency, however, sometimes only one of them is significantly detected. Therefore we show the frequency-frequency relations for these two components together. Figures~\ref{fig:wkrel} and \ref{fig:pbkrel} show that most of the frequencies we measured in the \emph{different states} fall on the pre-existing WK and PBK relations, although the LS and hard VHS show a larger dispersion in frequency, especially in $\nu_{LF}$. In most of the states there are correlated changes in the frequencies and generally the frequencies of the components are higher in the hard and soft VHS, hard IMS and HS than in the LS. 

\begin{figure}
\epsscale{1.0}
\plotone{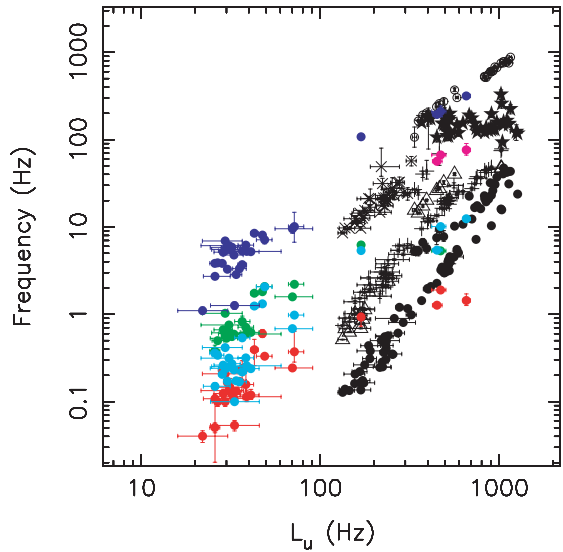}
\caption[]{Correlations between the characteristic frequencies of the power spectral components as a function of $\nu_{u}$. In black the original points for the Atoll sources, Low luminosity bursters, XTE~J1814--338, SAX~J1808.4--3658, XTE~J1751--305 and XTE~J0929--314 taken from  \cite{straaten05} and the measurements for XTE~J1807--294 taken from \cite{manu05}. The $\nu_{u}$ and $\nu_{\ell}$ values for SAX~J1808.4--3658 and XTE~J0929--314 have been multiplied with a factor 1.5 following \cite{straaten05} and those for XTE~J1807--294 with a factor 1.59 following \cite{manu05}. The solid dots represent L$_{b}$, the pluses L$_{h}$, the crosses L$_{low}$, the stars L$_{hHz}$, the triangles L$_{LF}$ and the open circles L$_{\ell}$. The colored dots represent the measurements for the black hole sources in our sample: in red L$_{b}$, in light-blue L$_{LF}$, in green L$_{h}$, in purple L$_{hHz}$ and in dark-blue L$_{\ell}$.}
\label{fig:stevemarc}
\end{figure}

The L$_{u}$ component is not covered by the WK and PBK relations presented in Figs.~\ref{fig:wkrel} and \ref{fig:pbkrel}, therefore in Fig.~\ref{fig:stevemarc} we present the relation of L$_{u}$ with all the other frequency components for our black hole measurements and those for atoll sources, Low luminosity bursters, XTE~J1814--338, SAX~J1808.4--3658, XTE~J1751--305, XTE~J0929--314 and XTE~J1807--294 as given by \cite{straaten05} and \cite{manu05}. It is clear from Fig.~\ref{fig:stevemarc} that the black hole components L$_{b}$, L$_{LF}$, L$_{h}$ and L$_{\ell}$ all follow more or less similar trends with respect to L$_{u}$ which, however, differ from the original relations for the neutron star sources \citep{straaten05,manu05}. Power law fits to the relations for the black holes presented in Fig.~\ref{fig:stevemarc} give significantly different values for the power law indices compared to the values given by \cite{straaten05}. But, when converting these indices to indices for power laws representing the WK and PBK relation for both our measurements for the black holes and those for the neutron stars given by \cite{straaten05} leads to similar results: while the relation of L$_{u}$ with all the other frequencies is systematically different between black holes and neutron stars, the relations between the LF QPO and L$_{b}$ and between the LF QPO and L$_{\ell}$ are very similar. Furthermore, Fig.~\ref{fig:stevemarc} also shows that all the L$_{u}$ points for the black holes are consistently too \emph{low} compared to the neutron star measurements. The same is also true for the L$_{hHz}$ features in Fig.~\ref{fig:stevemarc}: although their frequencies seem to be more or less independent of L$_{u}$, similarly to the hecto-Hz features found in neutron stars \citep[e.g.][]{straaten05}, they are found at lower frequencies compared to the those neutron star hecto-Hz features. However, we only detect a L$_{hHz}$ in combination with L$_{u}$ on three occasions in our data sample (in the soft VHS) so unfortunately we lack statistics to come to any firm conclusions regarding the behavior of this component.

So, based on the WK and PBK relations (Figs.~\ref{fig:wkrel} and \ref{fig:pbkrel}) one would identify most power spectral components in black holes and neutron stars as being the same (despite frequency and amplitude differences), but the component identified as L$_{u}$ as \emph{different} in black holes and neutron stars. 

\section{Discussion}
\label{sec:discbhns}

In the previous sections we have suggested an identification for the power spectral components observed in the various canonical black hole states. It is based on the changes observed in the frequencies and amplitudes of the power spectral components during canonical state changes that allow us to trace the components across states. An important new ingredient that we used in interpreting the power spectra is the occurrence of blends between components both at high and low frequency. We demonstrated that the occurrence of such blends is plausible based on the systematic behavior of the components that is observed when they are detected separately.

The basic idea behind the identification, that there are smooth transitions between different power spectral states which allow for individual components to be traced, is also supported by the work of \cite{nowak00}, \cite{kalemci03,kalemci05} and \cite{bell05b}. In particular \cite{bell05b} find that during the 2002/2003 outburst of GX~339--4 a LF QPO in the hard VHS can be linked to a QPO observed in the LS, and that the overall power spectral characteristics of the hard IMS can be seen as a continuation of those in the LS; all the power spectral frequencies increase when going from spectrally hard to soft, and decrease when going from soft to hard. Furthermore, they find sharp transitions to and from the soft VHS similar to the ones we find in our ensemble of sources. Although \cite{nowak00} give a similar identification for the broad "high frequency" components in the LS spectra of GX~339-4 and Cyg~X--1 as the one we use (Fig.~\ref{fig:identify}e) and extend to the other states, further examination of the high frequency components is required.

\subsection{The identification of the high-frequency components}
\label{sec:caveats}

The possible blends and the presence of sharp high frequency QPOs in addition to the broad features reported here, imply several caveats regarding the identification of the high frequency components L$_{hHz}$, L$_\ell$ and L$_{u}$ across the different canonical states.

The possibility of observing a blend between L$_\ell$ and L$_{u}$, was discussed in Sect.~\ref{sec:ident}. In most cases it may be a lack of statistics that prevents us from significantly detecting two separate features; either a second feature above L$_\ell$ is not significantly detected (but occasionally has a relatively high upper limit, see Sect.~\ref{sec:ident}) or L$_\ell$ is found to be relatively broad (typical Q values of less than 2), suggestive of a blend. For black hole sources in the original PBK relation the strongest LF QPO is identified with the horizontal branch oscillation (HBO) found in Z sources, based on the similar characteristic relations with both count rate and with the break frequency \citep[][]{pbk99}. The high frequency QPOs, broad and narrow, observed in black holes at that time did not share any characteristics with the kHz QPOs found in the neutron star sources. But, predominantly based on the similarity with the neutron star source Cir~X--1 \cite{pbk99} nevertheless identified these features with the lower kHz QPOs. \cite{bpk02} used the same identification as \cite{pbk99}, and neither of them address the possibility that the PBK relation contains a blend of L$_{\ell}$ and L$_{u}$ components all designated L$_{\ell}$. Our identification of the components in the LS power spectrum is based on that given by \cite{bpk02} in which the two broad high frequency features called  L$_{\ell}$ and L$_{u}$ (see Sect.~\ref{sec:lshvhshims}) are tentatively identified with the lower and upper kHz QPOs found in neutron stars and where in case of only one high frequency feature the designation L$_{\ell}$ is always used. For this reason the high frequency feature above the low frequency QPO called L$_{\ell}$ by both \cite{pbk99} and \cite{bpk02} is physically the same as the blended feature we identify here. So, we can safely and consistently compare our results with those of \cite{pbk99} and \cite{bpk02} when we designate the high frequency feature as L$_{\ell}$.

The match of the frequencies of our identified components to the pre-existing frequency-frequency relations (Sect.~\ref{sec:ffrel}) provides an argument against other identifications that fit less well. As an example of an alternative identification, instead of identifying a single high frequency feature as a blend between L$_{\ell}$ and L$_{u}$, for the sequence of observations presented in Figs.~\ref{fig:overgang} and \ref{fig:overgang2} another possibility would be that sometimes we see a blend between L$_{h}$ and L$_{\ell}$. For instance, in Figs.~\ref{fig:overgang}a--c we can identify the feature now classified as L$_{h}$ as a blend of L$_{h}$ and L$_{\ell}$ and the feature detected above that (now designated L$_{\ell}$) as L$_{u}$. In that case we only detect L$_{h}$ and L$_{\ell}$ as separate features in Figs.~\ref{fig:overgang}d; i.e. only in power spectra that have significant detections of L$_{u}$. However, this identification is \emph{not} supported by the PBK relation: if we take for $\nu_{\ell}$ the frequency of the blended features L$_{h}$ and L$_{\ell}$, then we find that $\nu_{\ell}$ is systematically too low. We also checked other alternative identifications against the pre-existing frequency-frequency relations and did not find one that showed a better match than the one presented here.

\input{tab1.tex}

The situation is more complicated when three features are detected at high frequency, as occasionally occurs in the soft VHS. An example of such a situation is presented in Fig.~\ref{fig:3hfqpos}; no low frequency broad noise component, L$_{h}$, is found, but an extra feature is detected at higher frequencies which we designated L$_{hHz}$. We identify this feature to be different from the L$_{h}$ component based on their different characteristics: the L$_{h}$ in the hard VHS has on average a frequency of 5.7$\pm{2.4}$ Hz, a Q value of 0.5$\pm{0.1}$ and an rms amplitude of 5.0$\pm{0.7}$\% while the L$_{hHz}$ has on average a frequency of 40.7$\pm{5.0}$ Hz, a Q value of 0.9$\pm{1.3}$ and an rms amplitude of 1.7$\pm{0.3}$\%. Because we only sporadically detect the hecto-Hz features, we cannot trace them across the states; L$_{hHz}$ is found in between L$_{LF}$ and L$_\ell$ and the feature above L$_\ell$ is identified with L$_{u}$. Table~\ref{t:qpochar} gives the frequency range, and the range of Q and rms values for the hecto-Hz QPOs (as well as all the other components) detected in our sample. Note that \cite{straaten02} suggested L$_{hHz}$ in black holes to be similar to the hecto-Hz features detected in the neutron star sources. Features similar to L$_{hHz}$ found here have been reported for other black hole sources, for instance in XTE~J1650-500 \citep{homan03a, kalemci03} and 4U~1630--47 \citep{klein05}. However, as concluded in Sect.~\ref{sec:ffrel} while the hecto-Hz features in the neutron star sources all have frequencies close to $\sim150$ Hz \citep[e.g.][]{straaten02}, the L$_{hHz}$ in black holes are found at lower frequencies between $\sim20$--$80$ Hz (Table~\ref{t:qpochar} and Fig.~\ref{fig:stevemarc}).

The high frequency features reported here are usually broad and it is not clear how they are related to the sharp high frequency QPOs (Q values of 2 and larger) in black holes reported in the literature \cite[e.g.][]{stroh01a,rem02c,homan05a,homan03a,miller01}. Although occasionally we detect sharp high frequency features at similar frequencies as reported by previous authors, the single trial significances were relatively low. This is most likely related to the fact that special selections on time or energy are necessary \citep[e.g.][]{stroh01a} for the significant detection of the black hole kHz QPOs, which were not made in the analysis presented here. The detailed analysis of the behavior of the high frequency features (kHz QPOs) is an important one, but is restricted by the limited statistics at those frequencies for the black hole power spectra: most of these features are only barely detected above the noise level, and up to now we only have a handful of significant detections. A more comprehensive study of the behavior of these features is beyond the scope of this paper.

\subsection{Gradual canonical state changes and implications regarding the identification of the black hole power spectral components}
\label{sec:gradual}

The behavior of transients and persistent sources has for many years been described in terms of fixed canonical states, both for black holes and neutron stars (Sect.~\ref{sec:introbhns}). Although this rather rigid scenario has provided a reasonable description of black hole sources, the increasingly detailed coverage of the transient sources with RXTE over the last few years has provided a more complete picture. There has been some debate on the classification of the black hole states and on the descriptors that define a state, however, it generally comes down to a combination of energy spectral and power spectral behavior. The recent outbursts of for instance XTE~J1550--564, GX~339--4 and GRO~J1655-40, have intense coverage especially during the rise and the decay of the outbursts filling in the "gaps" in the CD and HIDs left by the more patchy observations of the past. They present the source (and its power spectra) as exhibiting a continuum of spectral and temporal behavior that can be usefully parametrized through their location in the HID (see Fig.~\ref{fig:hidstatesBH}) which is closely coupled to the power spectral shape. Observations like these present a picture in which the various states represent repeated occurrences of the source with clearly different energy- and power spectral behavior. Between some of these states smooth transitions occur during which the power spectra gradually change while other transitions are more abrupt, as shown in Sect.~\ref{sec:ident} \citep[but see also for instance][]{rem02a,kalemci03}. Whether this is due to to just shorter timescales for the state change or true discontinuities in state is not yet clear.

Gradual changes in the power spectra have been reported before for a number of sources \citep[e.g.][]{homan01,bell05b,kalemci03} and in Sect.~\ref{sec:ident} we use this fact to trace the power spectral components across the different states. These gradual state transitions are ordinarily characterized by frequency (and amplitude) shifts when going from the spectrally hard state (LS) to the softer states (hard VHS/soft VHS/hard IMS) and vice versa, and generally frequencies decrease and fractional amplitudes increase when the source gets spectrally harder. Although the aperiodic variability often shows gradual changes, the state change is often marked by a more abrupt change in the spectral hardness \citep[see also][]{kalemci04}. However, in Sect.~\ref{sec:ident} we have discussed transitions between the hard and soft VHS that are characterized by a sudden disappearance of a noise component (L$_{h}$) in the power spectra (while most components suddenly become weaker). More dramatic changes in the power spectra occur for the HS -- soft/hard VHS and HS -- hard IMS transitions; on a relatively short time scale (hours to days) the power spectra change completely and so far no gradual change has been observed for these transitions. So, there seems to be a difference between gradual and more abrupt state transitions, where the gradual ones occur between the LS and hard IMS/VHS and the abrupt ones occur between the hard and the soft VHS and between the HS and either the hard/soft VHS or the hard IMS.

Regardless of whether the state transition is gradual or more abrupt, the identifications proposed here imply that the \emph{same} components are found before and after a transition. This is also the case for state transitions across the jet-line, which has been associated with the switching on and off ("quenching") of the radio jet \citep[see Fig.~\ref{fig:hidstatesBH} and][]{fender04,kording06}. In the current picture \citep[for a review, see][]{fender06} the radio : X-ray correlation is such that spectrally hard states (right-hand side of the jet-line) are related to radio emission. In the softer states (left-hand side of the jet-line) the radio emission is quenched and furthermore, when crossing the jet-line from right to left a large ejection event is predicted. The large change in the radio behavior when crossing the jet-line at high X-ray fluxes (see Fig.~\ref{fig:hidstatesBH}) is \emph{not} associated with a similar large change in the X-ray behavior: this transition across the jet-line is associated with the hard VHS to soft VHS transition which is characterized mainly by a disappearing noise component (L$_{h}$) while most of the other components are found on either side of the jet-line (compare Figs.~\ref{fig:identify}b and c). The largest change in the X-ray behavior occurs when the source makes the transition to the HS and the strength of all power spectral components decreases dramatically. Furthermore, while the hard VHS at high flux intervals is associated with radio emission (quasi-continuous jet), the hard IMS has almost identical power spectra (see Figs.~\ref{fig:identify}c and d) but has no radio emission. In fact, in cases when ample radio coverage is available the radio is only switched on when the transition to the LS occurs \citep[e.g.][]{homan05b}. Assuming that the outer-jet, located a distance of $\sim10^{5}$ Schwarzschild radii (for cm wavelengths) away from the compact object, is the source of the radio emission \cite[][]{sera05} the correlated radio : X-ray behavior clearly shows that whatever physical structures are responsible for the production of the X-ray variability they are not strongly related to the radio jet. Hence, the QPOs and noise components are not generated in the outer-jet, instead, presumably originate in the disk or corona (which could still form the base of the jet).

The association between some variability components identified as being the same in neutron star and black holes suggest that they originate from the same physical process. These processes cannot then be related to characteristics unique to either type of object, such as the material surface or magnetic field of the neutron star or to the horizon or extreme frame-dragging of a black hole. Instead, they are more likely associated to processes that take place in common structures such as disk, corona or jet (but see above). Another possibility is that the mechanisms producing the different variability components are \emph{different}, but that common processes in black hole and neutron star systems are responsible for creating the \emph{same correlation} between variability types. Again, in that respect the common processes are most likely related to the accretion flow \emph{around} the compact objects and not to unique physical characteristics of black holes or neutron stars themselves. It is for the first time that the two major frequency-frequency relations (PBK, WK; Figs. ~\ref{fig:wkrel} and \ref{fig:pbkrel}) known to hold for both neutron star and black hole systems in \emph{some} states are shown to hold for \emph{all} black hole states (Sect.~\ref{sec:ffrel}). Apparently the changes in the accretion flow that are supposed to exist between the different canonical states do not affect the correlation between some of the power spectral components, suggestive of a common process responsible for those correlations. However, as presented in Table~\ref{t:qpochar} there are clear differences between components interpreted as being the same. Besides the clear differences in frequency, we observe sometimes large differences in Q and rms amplitude. In the next section we discuss that amplitude difference in some more detail. 

The third important frequency-frequency relation originally proposed by \cite[][Fig.~\ref{fig:stevemarc}]{straaten03} does show clear differences in the behavior of one component, L$_u$, dubbed the same in both black holes and neutron stars. In black holes L$_u$ is consistently found at lower frequencies and shows a different relation to all the other components, compared to the L$_u$ component found in the neutron stars. While the difference in frequency in black holes and neutron stars can perhaps be explained as a mass effect (see Sects.~\ref{sec:bhvsns} and \ref{sec:bhnsstates}), the different dependence on the other frequencies suggests additional differences in the mechanism producing these components in black holes and neutron stars.

It is interesting to note that Figs.~\ref{fig:wkrel} and \ref{fig:pbkrel} show that the lowest frequencies which the various correlated components reach in the hard state are similar for the neutron stars and the black holes. Although selection effects associated with low source fluxes may play a role here (low sensitivity for the lowest frequencies as these are reached at the lowest luminosities), this argues against the possibility to estimate compact object mass from LS observations of black holes to an accuracy that would allow to detect the factor 5-10 difference between neutron stars and black holes in LMXBs. The highest frequencies appear to be attained by these same correlated components in the hard IMS/VHS (the soft VHS frequencies are somewhat lower), and here we see a factor $\sim$5 or more difference, consistent with expected neutron star and black hole mass ratios. So the maximum frequencies reached in these states may be a more reliable mass indicator. Of course, as these frequencies are variable, estimating masses from them is anyway somewhat risky.

The link observed by \cite{bell05b} of their "type C" QPO in the hard IMS to a QPO observed in the LS agrees with our classification for the components in the hard VHS and IMS, which uses the transition to and from the LS. Further distinctions into type A, B and C have been proposed among the L$_{LF}$ QPOs \citep[see for instance][and references therein]{whk99,cas05}. Although we do not use that subdivision here, the QPO that we identify as L$_{LF}$ in the LS and hard IMS and VHS most likely corresponds to type C and that in the soft VHS to type B (Fig.~\ref{fig:identify}). These QPOs are often rather similar and the main difference between the hard and soft IMS power spectrum occurs in the noise components. Interestingly, the differences between type A, B and C QPOs reported \cite{bell05b} and others, do apparently not manifest themselves in the frequency-frequency relations: we find that all the L$_{LF}$ components across different states follow the PBK and WK relations. In this context we note again that rare features tend to be disregarded in our approach that focuses on characteristic behavior.

\subsection{A comparison between black holes and neutron stars}

\subsubsection{Black hole power vs. neutron star power}
\label{sec:bhvsns}

\begin{figure}
\epsscale{1.15}
\plotone{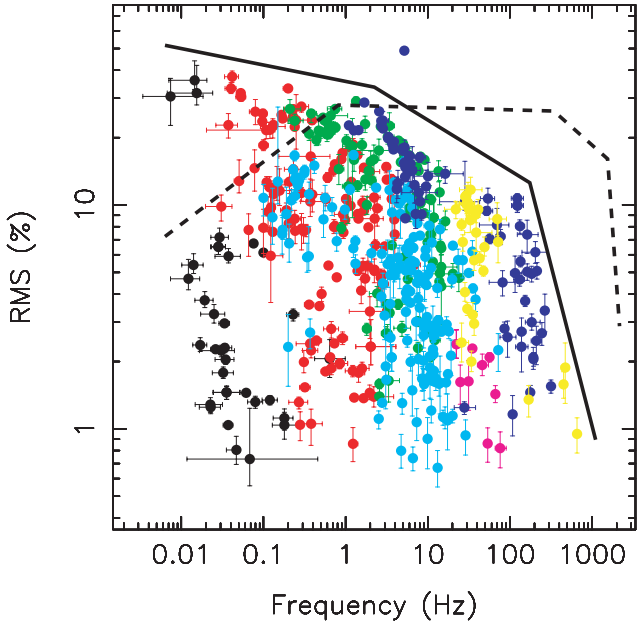}
\caption[]{RMS amplitude (in \%) vs. the frequency for a representative set of observations from the sample of black hole sources. The different power spectral componets are indicated with different colors: VLFN--Black, L$_{b}$--Red, L$_{LF}$--Light-blue, L$_{h}$--Green, L$_{hHz}$--Purple, L$_{\ell}$--Dark-blue and L$_{u}$--Yellow. Two contours are drawn, the line for the black hole points in this figure and a dashed one for the neutron star measurements from Fig.~\ref{fig:nsplot}.}
\label{fig:bhplot}
\end{figure}

\begin{figure}
\epsscale{1.15}
\plotone{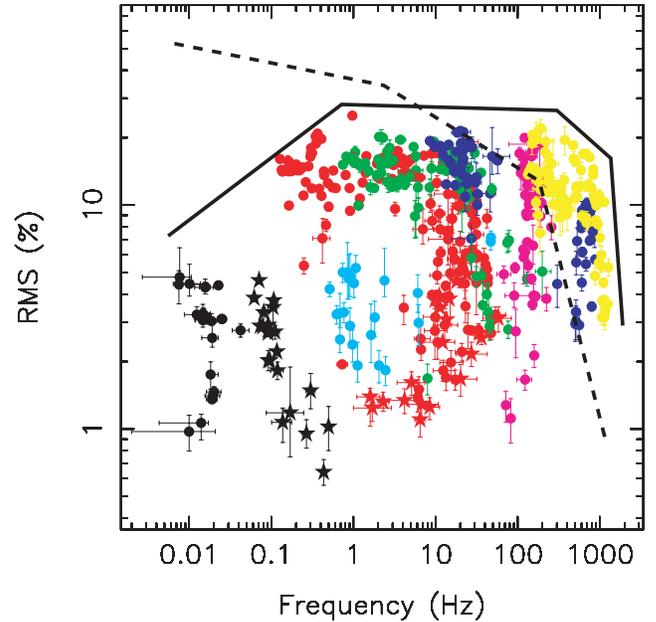}
\caption[]{RMS amplitude (in \%) vs. the frequency for a representative set of observations from the sample of neutron star sources suplemented by points taken from \cite{straaten02,straaten03,straaten05}. The different power spectral componets are indicated with different colors: VLFN--Black, L$_{b}$--Red, L$_{LF}$--Light-blue, L$_{h}$--Green, L$_{hHz}$--Purple, L$_{\ell}$--Dark-blue and L$_{u}$--Yellow. The black and red stars are the (high-) VLFN and L$_{b}$ components taken from \cite{reer04}. Two contours are drawn, the line for the neutron star points in this figure and a dashed one for the black hole measurements from Fig.~\ref{fig:bhplot}.}
\label{fig:nsplot}
\end{figure}

With our proposed variability component identifications we can now perform a first investigation into the origin of the finding of \cite{sun00} that the neutron stars show more power than black holes at high frequencies. In Figs.~\ref{fig:bhplot} and \ref{fig:nsplot} we plot the fractional rms amplitude of the different power spectral components vs. their $\nu_{max}$ frequency for black holes and neutron stars, respectively. These measurements are obtained from fits to the same representative set of 160 observations also used in Figs.~\ref{fig:wkrel} and \ref{fig:pbkrel}. For the neutron stars we use additional points taken from \cite{straaten02,straaten03,straaten05} and the measurements by \cite{reer04} for the atoll sources GX~9+9, GX~9+1 and GX~3+1 \footnote{Note that the measurements by \cite{reer04} were used since it provides more characteristics examples of UB and LB power spectra than were present in our sample}. \cite{reer04} fit the VLFN with a power law and therefore those measurements cannot be used here as no central frequency is provided. However, for those VLFN components they find that these contribute to the total power in the range below 0.1 Hz with fractional rms amplitudes in the range of 2--4\% rms, which is at a similar level as the VLFN measurements presented here (black points in Fig.~\ref{fig:nsplot}). Note that as all states are mixed together in Figs.~\ref{fig:bhplot} and \ref{fig:nsplot}, the usual order of frequencies between components is mixed up as well. Also, Figs.~\ref{fig:bhplot} and \ref{fig:nsplot} do not contain information pertaining to the frequency range over which the power of each component is distributed. Using for instance the FWHM of the features to indicate this for each data point complicates the figure but does not lead to different insights.

Figures~\ref{fig:bhplot} and \ref{fig:nsplot} both show the two contours representing the upper envelope of the $\nu_{max}$-rms distribution of black holes and neutron stars as black and dashed lines and vv., respectively, to allow for easy comparison between the two types of sources. The contours cross around $5$ Hz, with the neutron star features having larger rms amplitudes above that frequency, while the black hole features are stronger below it. Above $5$ Hz the strongest black hole features are weaker than those in neutron stars by up to a factor 6 (i.e., 36 in power) and have characteristic frequencies of up to a few 100 Hz, while neutron star features have frequencies up $\sim1000$ Hz.


\begin{figure}
\epsscale{1.0}
\plotone{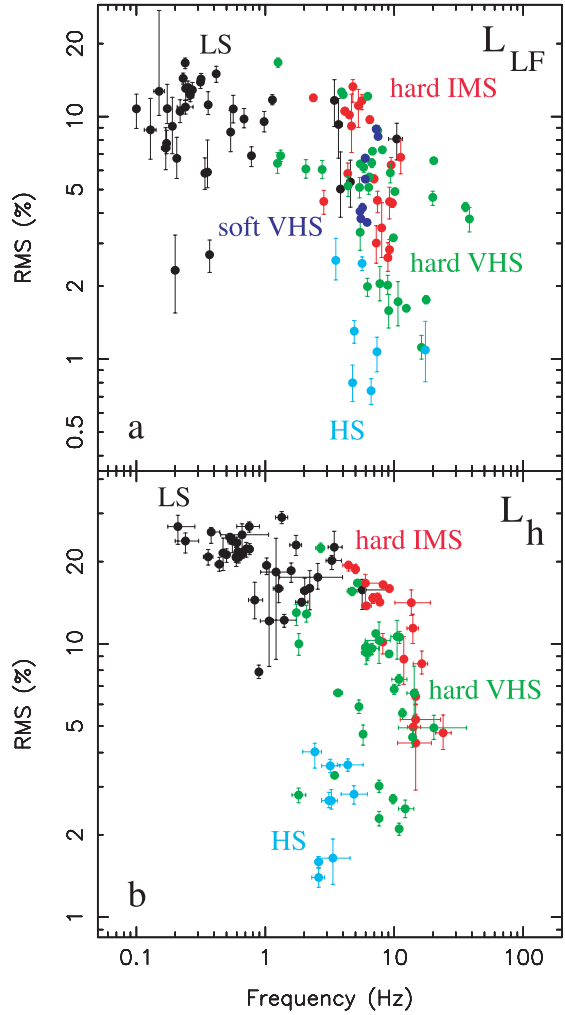}
\caption[]{Fractional RMS as a function of frequency for L$_{LF}$ (panel a) and L$_{h}$ (panel b) shown for the LS (black), HS (light-blue), hard IMS (red), hard VHS (green) and the soft VHS (dark-blue) observations (points similar to those in Fig.~\ref{fig:bhplot}). }
\label{fig:rmsllflh}
\end{figure}

\begin{figure}
\epsscale{1.0}
\plotone{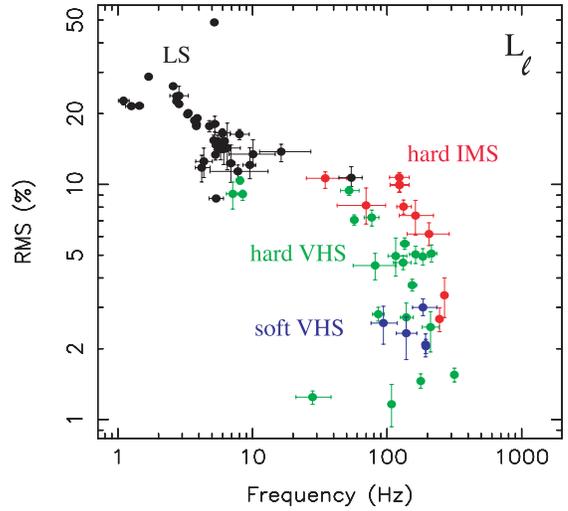}
\caption[]{Fractional RMS as a function of frequency for L$_{{\ell}}$ for the LS (black), hard IMS (red), hard VHS (green) and the soft VHS (dark-blue)observations (points similar to those in Fig.~\ref{fig:bhplot}). }
\label{fig:rmsll}
\end{figure}

The general trend across about 5 decades in frequency, as presented in Figs.~\ref{fig:bhplot} and \ref{fig:nsplot}, is somewhat different for black holes and neutron stars: for neutron stars the frequency-rms distribution envelope is more flat than that for black holes, and the decrease in power is much steeper and occurs at higher frequencies, well above a few $100$ Hz. Furthermore, the frequency-rms distribution reveals the widely acknowledged observation that the power spectral components tend to be stronger and found at lower frequencies in the spectrally hard states compared to the softer states (see also Figs.~\ref{fig:rmsllflh} and \ref{fig:rmsll}).

Using Figs.~\ref{fig:bhplot} and \ref{fig:nsplot} we can identify which components are responsible for the broad-band power spectral differences between black hole and neutron star X-ray binaries. At frequencies above $\sim5$ Hz the neutron stars dominate due to their stronger L$_{hHz}$, L$_{\ell}$ and L$_{u}$ features. The color-coding also shows that similar power spectral features tend to be found at higher frequencies for neutron star sources. So, the differences in power between neutron stars and black holes at high frequencies reported by \cite{sun00} can all be attributed to differences in frequency and strength of these high frequency features, and do not require the \emph{absence} of any components from black holes that are present in neutron stars. Note however that, as explained in Sect.~\ref{sec:gradual} L$_{u}$ is found to be different in both types of systems. Assuming that the frequencies are inversely related to the mass of the accreting object \footnote{In the simplest case all frequencies are Keplerian at radii that scale with the Schwarzschild radius, r=6GM/c$^{2}$, for which the frequency is $\nu_{K}=\sqrt((G M)/r^{3})$ and hence: $\nu_{K}\sim\frac{1}{M}$}, the observed frequency difference between neutron stars and black holes might be a manifestation of the mass difference between them. Furthermore, the differences in amplitude might be related to the presence of a solid surface in neutron stars, absent in black holes, which enhances the effect of any inhomogeneities in the accretion flow by affecting not only the integrated disk emission but also the boundary layer/surface emission at the time the inhomogeneities accrete.

At frequencies below $\sim5$ Hz, the black holes are up to 8 times stronger then the neutron stars. This is predominantly caused by the strong L$_{b}$ component in the black hole LS spectra, see Fig.~\ref{fig:reeks}, and only rarely by a stronger VLFN (i.e. a component below L$_{b}$). Note that recently \cite{manu07} have shown that the accreting millisecond pulsar IGR~J$00291+5934$ shows band strong limited noise of $\sim12$--$40$\% (rms) in the frequency range between $\sim0.01$ and $0.05$ Hz, a behavior that is very unusual for this type of source. In fact, these high fractional rms amplitudes at low frequencies makes IGR~J$00291+5934$ very similar to black hole sources, see Figs.~\ref{fig:bhplot} and \ref{fig:nsplot}, as already pointed out by \cite{manu07}.

Figures~\ref{fig:rmsllflh} and \ref{fig:rmsll} show the behavior for the black hole L$_{LF}$, L$_{h}$ and L$_{\ell}$ components but now separated by source state. From these figures it is clear that the rms decreases only above a certain frequency, clearly related to the hard to soft transition transition, but only with little additional frequency change. For the other components in the black hole power spectra and for those in the neutron star power spectra similar trends are observed, although sometimes less clear. 

Finally, Fig.~\ref{fig:rmsll} clearly shows a bimodal behavior for the black hole L$_{\ell}$ component. The LS is characterized by L$_{\ell}$'s at relatively low frequencies, $\sim1$--$20$ Hz, with Q values between 0 and 0.6 and rms amplitudes between 10--50\%. On the other hand, the L$_{\ell}$ components in the hard-, soft VHS and hard IMS are found between $\sim10$--$300$ Hz with Q values between 0 and 6 and rms amplitudes of 1--10\%. This behaviour is very similar to what was found for the high frequency QPOs in the atoll sources. As  mentioned before, \cite{straaten05} distinguished two types of lower kHz QPO features in the neutron star atoll sources: L$_{\ell ow}$ for broad, relatively strong features at frequencies below $\sim100$ Hz and L$_{\ell}$ for sharp (Q$>2$) features at kHz frequencies, see Table~\ref{t:qpochar}. Because we identify the broad high frequency feature in the hard-, soft VHS and hard IMS as a possible blend of two features, we have to be careful when comparing these measurements with those in the LS where we do detect L$_{\ell}$ and L$_{u}$ as seperate features. For instance, we are very likely overestimating the width (or underestimating the Q) of the high frequency features (designated  L$_{\ell}$). Therefore, for the black hole power spectra we have not made the distinction between L$_{\ell}$ and L$_{\ell ow}$.

\subsubsection{black hole and neutron star states}
\label{sec:bhnsstates}

In the previous sections we have shown that, based on the identification of all the components in the black hole power spectra and using the known frequency-frequency relations, we can identify most of the variability components as being the same in black hole and neutron star X-ray binaries -- it is only L$_{u}$ that seem to be different. From Figs.~\ref{fig:hidstatesBH} and \ref{fig:hidstatesNS} it is clear that the black hole LS and neutron star EIS have similar relative positions in the HID. Also, as a basis for the identification we use the large similarities between the black hole LS and neutron star EIS power spectra: the overall shape is very similar, see Figs.~\ref{fig:reeks} and \ref{fig:identify} and when we shift the LS power spectrum by about a factor 5 in frequency such that L$_{b}$ in the LS is at the same frequency as the same component in the EIS, most other components coincide in frequency (but still can have considerable differences in amplitude). Obvious differences between the LS and EIS power spectra are the lack of LF QPOs but the much stronger high frequency features in the EIS while at lower frequencies the LS power spectra are stronger, similar to what we already concluded from Figs.~\ref{fig:bhplot} and \ref{fig:nsplot}. So, how similar are the power spectra of the other black hole and neutron star states?

Based on the relative position in Figs.~\ref{fig:hidstatesBH} and \ref{fig:hidstatesNS} it is reasonable to compare the black hole hard- and soft VHS and the hard IMS with the neutron star IS and perhaps also the LLB. The power spectra in Figs.~\ref{fig:reeks} and \ref{fig:identify} show that the IS and LLB lack the strong LF QPOs that dominate the black hole power spectra around $\sim10$ Hz, but have much stronger L$_{hHz}$, L$_{\ell}$ and L$_{u}$ features. Interestingly, a similar disappearance of the noise component L$_{h}$ from the hard- to the soft VHS is also seen between the IS and banana states. No shift in frequency to the black hole hard- and soft VHS and the hard IMS power spectra, such as required in the case of the LS and EIS, can be envisaged that results in most of the power spectral features to coincide. Although, a shift with a factor $\sim5$ does improve the match between the broad-band power spectral shapes.

Based on the relative position in the HID (Figs.~\ref{fig:hidstatesBH}, \ref{fig:hidstatesNS}) it would seem reasonable to compare the black hole HS with the neutron star LB and UB. However, the power spectra in Figs.~\ref{fig:reeks} and \ref{fig:identify} show no obvious similarities and no shift in frequency can be found that would consistently match power spectral components, or even improve the match between the broad-band power spectral shapes.

If the shift with frequency applied to the black hole power spectra with respect to the neutron star ones improves the match between the broad-band power spectral shapes and coincides some of the power spectral components that are dubbed the same in at least some of the canonical states, then the difference in frequency by a factor $\sim5$ can be a mass effect (see Sect.~\ref{sec:bhvsns}). However, a preliminary analysis has not provided conclusive results.

\section{Summary and Conclusions}

From a uniform analysis of a large (8.5 Ms) data set of RXTE/PCA observations of LMXBs we present for the first time a complete identification of all the components in the power spectra across the different canonical black hole states. It is based on the identification of the components in the LS that has a strong analogy with that in the neutron star EIS, and uses the observed frequency changes of the components as the source changes from one state to another. The identification is supported by the known frequency-frequency relations (PBK and WK) which are shown to hold for \emph{all} black hole canonical states. We find that:

\begin{itemize}
\item{for black holes the \emph{same} component is found in the different canonical states at a different frequency, but always following a fixed relation to the other frequencies;}
\item{L$_{b}$, L$_{LF}$, L$_{h}$ and L$_{\ell}$ in black hole power spectra are strongly related to the components that are dubbed the \emph{same} in neutron star power spectra but that are found at different frequencies again following a fixed relation to the other frequencies;}
\item{L$_{u}$ components in black holes seem to be different to the ones found in neutron stars: they are found at lower frequencies and have a different relation with all the other variability components;}
\end{itemize}

The observation of similar variability components in black holes and neutron stars can be explained in two ways. Either the variability originates from the same physical processes that take place in the accretion flow around the compact objects, or the origin of the variability might be different, but there is one generic mechanism that causes the correlations between the frequencies to be the same. Likewise, the same conclusions can be drawn from the identification of the \emph{same} frequency components across \emph{all} canonical black hole states: either the same physical processes occur throughout all canonical black hole states and state-specific characteristics are responsible for the observed differences in frequency and strength, or the underlying processes are different from state to state and one common mechanism is responsible for the frequency correlations. In any case, the common mechanism that either is responsible for only the correlations or determines all the characteristics of the variability cannot depend on unique source (or state) characteristics, such as the magnetic field or the surface of the neutron star or the spin of the black hole but instead must depend on the general properties of the accretion flow around compact objects.

The different behavior of L$_u$ in the black holes compared to the same component in neutron stars manifests itself in a lower frequency and a different relation to all the other frequencies. While the difference in frequency can perhaps be explained as a mass effect, the different dependence on the other frequencies suggests additional differences in the mechanism producing these components in black holes and neutron stars.

There are also caveats regarding the proposed identification one of which is related to the features detected at high frequencies (above $\sim100$ Hz) in the black hole power spectra. These features are relatively broad compared to the high frequency QPOs in black holes occasionally reported in the literature, and are therefore regarded as different. Furthermore, it cannot be ruled out that on some occasions (in the hard- and soft VHS and hard IMS) we see a blend of two high frequency features, L$_{\ell}$ and L$_{u}$: two sharper high frequency features cannot be individually resolved as they are closely spaced in frequency. In compliance with \cite{pbk99} we classified these blended features L$_{\ell}$. Also, at lower frequencies blends occur, especially between two relatively broad BLN components L$_{b}$ and L$_{h}$. 

The suggested identification is based on the observed gradual transitions between some power spectral states; while the transitions between the LS and hard IMS/VHS are found to be gradual, the ones between the soft states (hard--soft VHS, HS--hard/soft VHS / hard IMS) are found to be more abrupt. Whether this observed fact is due to shorter timescales for the state change or true discontinuities in state is not yet clear. However, the fact remains that the same power spectral components are found before and after the transition and in particular this is also true for power spectra on either sides of the so called jet-line (which marks the switching on and off of the radio jet). As also the association with radio emission of the hard VHS and hard IMS which occur at different flux levels but have very similar power spectra is ambiguous, it is suggested that the X-ray variability is not originating from the outer-jet, but most likely produced in either the disk or the corona (base of the jet).

The suggested classification also allow us to directly compare the strength of the variability between black holes and neutron stars. We find that:
\begin{itemize}
\item{at frequencies below $\sim5$ Hz black holes can show up to 8 times more power compared to neutron stars, predominantly due to L$_{b}$;}
\item{above $\sim5$ Hz black holes show up to a factor 6 less power, by means of weaker L$_{hHz}$, L$_{\ell}$ and L$_{u}$ features which can be related to the mass difference between the black holes and neutron stars;}
\item{the differences in power between neutron stars and black holes at high frequencies reported by \cite{sun00} can all be attributed to differences in both frequency and strength of these high frequency features, and do \emph{not} require the \emph{absence} of any components from black holes that are present in neutron stars, but note that L$_{u}$ is suggested to be different in both type of systems;}
\item{in the spectrally hard states black holes and neutron stars reach similar low frequencies, while in the softer states the neutron stars can reach a factor 5-10 higher frequencies, suggesting that the latter is a more reliable mass indicator;}
\item{a shift in frequency by about a factor 5 applied to the black hole power spectra with respect to the neutron star ones improves the match between the broad-band power spectral shapes. Some of the power spectral components that are dubbed the same coincide in at least some of the canonical states (LS -- EIS; hard-,soft VHS, hard IMS -- IS, LLB), which might again be a mass effect.}
\end{itemize}

The suggested identification of the power spectral components across all canonical black hole states serves as a first handle when studying the behavior of these sources and when comparing them with neutron star sources. The fact that the same components are found across all states and even between different types of sources is highly suggestive for a common underlying physical mechanism and put heavy constrains on the accretion disc theories that try to explain the wealth of complicated behavior in both black hole and neutron star X-ray binaries.

\acknowledgements
MKW would like to thank Thomas Maccarone, Rob Fender, Sera Markoff and Rudy Wijnands for their useful discussions and suggestions. MK acknowledges the Aspen Center for Physics for hospitality and participants in the Aspen Summer Workshop 'Revealing Black Holes' for stimulating exchanges.

\bibliographystyle{aa}
\bibliography{kleinwolt}

\end{document}

%% file: tab1.tex
\begin{table}

\centering
\begin{tabular}{cccc}
\hline
\hline
\multicolumn{4}{c}{Black Hole}		\\
\hline
Component &Freq (Hz) &Q &RMS (\%) \\
\hline
L$_{b}$		&0.03--7.0	&0.0-5.0	&0.9--37.0   \\
L$_{LF}$	&0.1--38.0	&1.1--31.0	&8.2--16.0   \\
L$_{h}$		&5.0--25.0	&0.0--2.0	&1.4--30.0    \\
L$_{hHz}$	&20.0--80.0	&0.4--5.0	&0.8--2.5     \\
L$_{\ell}$	&1.0--320.0	&0.0--6.7	&1.2--49.0    \\
L$_{u}$		&22.0--470.0	&0.001--8.0	&1.35--12.0   \\
\hline
\\[0.05cm]
\hline
\hline
\multicolumn{4}{c}{Neutron star}		\\
\hline
Component &Freq (Hz) &Q &RMS (\%) \\
\hline
L$_{b}$		&0.2--55	   &0.0--4.0	   &1.7--25 \\
L$_{LF}$	&0.5--48.0	   &2.3--11.0	   &1.9--7.2 \\
L$_{h}$		&0.7--200.0	   &0.0--6.2	   &1.7--20.0 \\
L$_{hHz}$	&68.0--330.0	   &0.0--8.9	   &1.1--20.0 \\
L$_{\ell ow}$	&8.5--57.0	   &0.0--1.47	   &7.0--21.0 \\
L$_{\ell}$	&300.0--880.0	   &4.0--50.0	   &3.0--10.0 \\
L$_{u}$		&134--1300.0	   &0.08--13.0     &3.0--22.0 \\
\hline

\end{tabular}
\caption{Characteristics of the black hole and neutron star variability components. Note that we give characteristic ranges for frequencies, Q and rms amplitudes across the different states. Note that, here we have made a subdivision in L$_{\ell ow}$ and L$_{\ell}$, with the latter corresponding to the spectrally hard neutron star EIS and black hole LS. }
\label{t:qpochar}
\end{table}